\documentstyle[sprocl,axodraw,cite]{article}

\makeatletter
\def\@cite#1#2{{$\!{[#1]}$\if@tempswa\typeout
        {IJCGA warning: optional citation argument
        ignored: `#2'} \fi}}
\makeatother

\makeatletter
\def\subsection{\@startsection{subsection}{2}{\z@}{-2.5ex plus -1ex
minus -.2ex}{1.5ex plus .2ex}{\it }}
\makeatother

\def\beq{\begin{equation}}
\def\eeq{\end{equation}}
\def\beqar{\begin{eqnarray}}
\def\eeqar{\end{eqnarray}}
\def\nn{\nonumber}

\def\Ga{\Gamma}
\def\de{\delta}

\def\refeq#1{\mbox{(\ref{#1})}}
\def\reffi#1{\mbox{Fig.~\ref{#1}}}
\def\reffis#1{\mbox{Figs.~\ref{#1}}}

\def\refse#1{\mbox{Sect.~\ref{#1}}}

\def\citere#1{\mbox{Ref.~\cite{#1}}}
\def\citeres#1{\mbox{Refs.~\cite{#1}}}

\newcommand{\GeV}{\unskip\,\mathrm{GeV}}
\newcommand{\MeV}{\unskip\,\mathrm{MeV}}

\renewcommand{\O}{{\cal O}}

\def\mathswitchr#1{\relax\ifmmode{\mathrm{#1}}\else$\mathrm{#1}$\fi}

\newcommand{\PV}{\mathswitchr V}
\newcommand{\PW}{\mathswitchr W}
\newcommand{\PZ}{\mathswitchr Z}

\newcommand{\PH}{\mathswitchr H}
\newcommand{\Pe}{\mathswitchr e}

\newcommand{\Pd}{\mathswitchr d}
\newcommand{\Pu}{\mathswitchr u}

\newcommand{\Pt}{\mathswitchr t}

\newcommand{\Pep}{\mathswitchr {e^+}}
\newcommand{\Pem}{\mathswitchr {e^-}}
\newcommand{\Pepm}{\mathswitchr {e^\pm}}

\newcommand{\PWp}{\mathswitchr {W^+}}
\newcommand{\PWm}{\mathswitchr {W^-}}

\newcommand{\PWpm}{\mathswitchr {W^\pm}}

\def\mathswitch#1{\relax\ifmmode#1\else$#1$\fi}

\newcommand{\MV}{\mathswitch {M_\PV}}
\newcommand{\MW}{\mathswitch {M_\PW}}

\newcommand{\MZ}{\mathswitch {M_\PZ}}
\newcommand{\MH}{\mathswitch {M_\PH}}
\newcommand{\Me}{\mathswitch {m_\Pe}}

\newcommand{\Mt}{\mathswitch {m_\Pt}}

\renewcommand{\ss}{\scriptscriptstyle}
\newcommand{\sw}{\mathswitch {s_{\ss\PW}}}
\newcommand{\cw}{\mathswitch {c_{\ss\PW}}}

\newcommand{\ri}{{\mathrm{i}}}
\newcommand{\rd}{{\mathrm{d}}}
\newcommand{\lsim}
{\;\raisebox{-.3em}{$\stackrel{\displaystyle <}{\sim}$}\;}

\begin{document}

\def\thefootnote{\fnsymbol{footnote}}
\setcounter{footnote}{0}
\null

\hfill CERN-TH/98-336 \\
\strut\hfill hep-ph/9811434
\vskip 3cm
\begin{center}
{\Large\bf
Radiative corrections to \\[.5em]
W-pair production in \boldmath{$\Pep\Pem$} annihilation%
\footnote{Contribution to  the proceedings of
{\it IVth International Symposium on Radiative Corrections, RADCOR 98},
8--12 September 1998, Barcelona, Spain.}
\par} \vskip 4em
{\Large
{\sc Stefan Dittmaier%
}\\[2ex]
{\large \it Theory Division, CERN\\
CH-1211 Geneva 23, Switzerland}\\[2ex]
}
\par \vskip 1em
\end{center}\par
\vskip .0cm
\vfill
{\bf Abstract:} \par
The status of precision calculations for the processes
$\Pep\Pem\to\PW\PW\to 4\,$fermions is reviewed, paying particular
attention to questions of gauge invariance and recent progress
concerning photonic radiative corrections.
\par
\vskip 1cm
\noindent
CERN-TH/98-336 \\
November 1998
\par
\null
\setcounter{page}{0}
\thispagestyle{empty}

\def\thefootnote{\alph{footnote}}
\setcounter{footnote}{0}

\title{
RADIATIVE CORRECTIONS TO \\[.5em]
W-PAIR PRODUCTION IN {\rm\bf e}$^+${\rm\bf e}$^-$ ANNIHILATION
}

\author{STEFAN DITTMAIER}

\address{CERN, Theory Division, CH-1211 Geneva 23, Switzerland \\
E-mail: Stefan.Dittmaier@cern.ch} 

\maketitle\abstracts{ 
The status of precision calculations for the processes
$\Pep\Pem\to\PW\PW\to 4\,$fermions is reviewed, paying particular
attention to questions of gauge invariance and recent progress
concerning photonic radiative corrections.
}

\section{Introduction}

At present, the most stringent tests of our understanding of the electroweak
interaction is obtained by confronting theoretical predictions with the
experimental precision data on the various Z-boson resonance observables
provided by LEP1 and the SLC, the Fermi constant $G_\mu$, the effective
electromagnetic coupling $\alpha(\MZ^2)$, and the masses $\MW$ and $\Mt$
of the W~boson and the top quark. The agreement between the
Standard-Model (SM) predictions and those experimental results, which is
at the impressive level of a few per mille, can be viewed as perfect
(see \citere{ho98} and references therein).
Despite this success, two cornerstones of the electroweak SM still
await direct empirical confirmation: the Higgs mechanism for the mass
generation and the detailed structure of the gauge-boson
self-interactions. The investigation of W-pair production in $\Pep\Pem$
annihilation, as observed at LEP2 and future $\Pep\Pem$ colliders,
yields important contributions to both aspects.

The experimental analysis of the reaction $\Pep\Pem\to\PWp\PWm$ allows 
for a precise determination of $\MW$. In the first half of 1998, the 
W-mass measurement at LEP2 \cite{mwlep} already reached a precision 
of $90\MeV$.
At LEP2 a final error of 30--40$\MeV$ 
\cite{lep2repWmass} is aimed at, which could be further reduced to about 
$15\MeV$ by observing W~pairs at a future $\Pep\Pem$ linear collider 
\cite{ecfa}. Such an improved knowledge of $\MW$ will strengthen the indirect
constraints on the mass $\MH$ of the SM Higgs boson, which are obtained
by fitting the SM parameters to the electroweak precision data.
In view of gauge-boson self-interactions, the process
$\Pep\Pem\to\PWp\PWm$ yields more direct information, since the
non-Abelian $\gamma$WW and ZWW couplings enter the perturbative 
predictions already in lowest order. It is expected that LEP2 
\cite{lep2repancoup} will test these triple gauge-boson couplings (TGCs) 
at the level of 10\% with respect to the SM coupling strength, and that 
a future linear collider \cite{ecfa} can even exceed the per-cent
level.

The desired information on the properties of the W~boson is
extracted from the cross section and the relevant angular and
invariant-mass distributions for the process $\Pep\Pem\to\PWp\PWm$.
More precisely, the W-boson mass is determined \cite{lep2repWmass}
by inspecting the total cross section near threshold, where it is most 
sensitive to $\MW$, and, sufficiently above threshold, by reconstructing the 
invariant mass of the W~boson from its decay products. In the LEP2
energy range, restrictive bounds on TGCs \cite{lep2repancoup} 
can only be obtained by considering various angular distributions, such 
as the one for the W-production angle. For higher energies, also the
total cross section becomes more and more sensitive to TGCs.
\looseness -1

The described experimental aims require the knowledge of the SM
predictions for the mentioned observables to a high precision.
For LEP2, the cross section of W-pair production should be
known within $\sim0.5\%$ \cite{lep2repWcs}; future linear colliders with
higher luminosity and energy set similar requirements. The theoretical
precision for the invariant-mass distribution of the W~bosons should, of
course, exceed the expected experimental accuray given above. 
To achieve this level of precision in predictions is a highly
non-trivial task, since the actually relevant reaction is the
four-fermion process $\Pep\Pem\to\PW\PW\to 4f$, in which the unstable 
W~bosons appear as resonances. The issue of gauge invariance requires 
particular attention, and the necessary inclusion of radiative corrections 
is not straightforward.
In this article, these sources of complications and their 
consequences for actual calculations are discussed, and special 
emphasis is laid on recent developments. 
More details can be found in other review articles 
\cite{lep2repWcs,be94,crad96,wwrevs} and references therein.

\section{The issue of gauge invariance}

\subsection{Finite gauge-boson widths and lowest-order predictions}
\label{se:widths}

\begin{sloppypar}
At and beyond the per-cent accuracy, gauge-boson resonances cannot be
treated as on-shell states in lowest-order calculations, since the impact 
of a finite decay width $\Ga_{\mathrm{V}}$ for a gauge boson V of mass 
$M_{\mathrm{V}}$ can be roughly estimated to 
$\Ga_{\mathrm{V}}/M_{\mathrm{V}}$, which is, for instance, $\sim3\%$ for the 
W~boson. Therefore, the full set of tree-level diagrams for a given fermionic 
final state has to be taken into account. For $\Pep\Pem\to\PW\PW\to 4f$ this
includes graphs with two resonant W-boson lines (``signal diagrams'') and
graphs with one or no W~resonance (``background diagrams''), leading to
the following structure of the amplitude \cite{lep2repWcs,be94,ae94}:
\looseness-1
\beqar
\label{eq:Mstruc}
{\cal M} =
\underbrace{\frac{R_{+-}(k_+^2,k_-^2)}{(k_+^2-\MW^2)(k_-^2-\MW^2)}
        }_{\mbox{doubly-resonant}}
+\underbrace{\frac{R_{+}(k_+^2,k_-^2)}{k_+^2-\MW^2}
            +\frac{R_{-}(k_+^2,k_-^2)}{k_-^2-\MW^2}
        }_{\mbox{singly-resonant}}
+\underbrace{N(k_+^2,k_-^2)}_{\mbox{non-resonant}}\!.
\nn\\*[-1em] \\*[-2.3em] \nn
\eeqar
\end{sloppypar}
\noindent
Gauge invariance implies that ${\cal M}$ is independent of
the gauge fixing used for calculating Feynman graphs (gauge-parameter
independence), and that gauge cancellations between different
contributions to ${\cal M}$ take place.
These gauge cancellations are ruled by Ward identities
(see e.g.\ \citeres{crad96,imflscheme,flscheme}), and their violation 
can completely destroy the consistency of predictions 
\cite{ku95,imflscheme,flscheme}. 

For a physical description of the W~resonances, the finite W~decay width
has to be introduced in the resonance poles.
However, since only the sum in
\refeq{eq:Mstruc}, but not the single contributions to ${\cal M}$, 
possesses the gauge-invariance properties, the simple replacement
\beq
\left[k^2-\MV^2\right]^{-1} \quad\to\quad
\left[k^2-\MV^2+\ri\MV\Ga_\PV(k^2)\right]^{-1}
\label{eq:naiveGV}
\eeq
\begin{sloppypar}
\noindent
in general violates gauge invariance. 
We describe in more detail some methods 
(see \citeres{lep2repWcs,flscheme,ae93,de98b} and references therein)
for introducing finite gauge-boson widths in lowest-order amplitudes.
\end{sloppypar}
\renewcommand{\labelenumi}{(\roman{enumi})}
\begin{enumerate}
\item 
The {\it fixed-width scheme} is nothing but the naive replacement 
\refeq{eq:naiveGV} for all gauge-boson propagators where 
$\Gamma_\PV(k^2)$ is the (constant) on-shell width $\Gamma_\PV$. 
In general, gauge
dependences are introcuded, and the Ward identities are violated. For
$\Pep\Pem\to 4f$, electromagnetic gauge invariance is maintained,
and the SU(2)-violating effects are suppressed by the factor
$\Gamma_\PW\MW/s$ in the high-energy limit \cite{flscheme}. 
\item 
The {\it running-width scheme} 
differs from the constant-width scheme only in the form of
the function $\Gamma_\PV(k^2)$ in \refeq{eq:naiveGV}, which is now
chosen as $\Gamma_\PV(k^2) = \Gamma_\PV \times \theta(k^2) k^2/\MV^2$.
Although the running width seems to describe the propagator in a more 
realistic way at first sight, for $\Pep\Pem\to 4f$ not even electromagnetic 
gauge invariance is retained anymore \cite{flscheme}. 
\item 
\begin{sloppypar}
The {\it complex-mass scheme} \cite{de98b}%
\footnote{Aspects of such a scheme have already been
discussed in \citere{st90} for the description of the Z-boson resonance
in $\Pep\Pem\to\PZ\to f\bar f$.} 
demands the consistent replacement $\MV^2\to\MV^2-\ri\MV\Gamma_\PV$
for the gauge-boson masses whenever a gauge-boson mass appears. In
particular, this leads to the complex weak mixing angle defined by
\beq
\cw^2 = 1-\sw^2 =
\frac{\MW^2-\ri\MW\Gamma_\PW}{\MZ^2-\ri\MZ\Gamma_\PZ},
\label{eq:csw}
\eeq
i.e.\ coupling constants become complex-valued.
As long as tree-level amplitudes are parametrized by a minimal set of
input parameters, algebraic relations between Feynman diagrams remain
unchanged so that gauge-parameter dependences still cancel, and
Ward identities still hold.
Despite the nice features of this scheme, it should be used with care,
as its full consistency is not yet clarified.
\end{sloppypar}
\item 
The {\it fermion-loop scheme} \cite{flscheme}
goes beyond a pure tree-level calculation by including and consistently
Dyson-summing all closed fermion loops in $\O(\alpha)$. This procedure
introduces the running tree-level width in gauge-boson
propagators via the imaginary parts of the fermion loops. 
Ward identities are not violated, since the
fermion-loop (as well as the tree-level) contributions to vertex functions 
obey the simple linear (also called ``naive'') Ward identities that are 
related to the original gauge invariance rather than to the more involved 
BRS invariance of the quantized theory.
Owing to the linearity of the crucial Ward identities for the vertex
functions, the fermion-loop scheme works both with the full fermion
loops and with the restriction to their imaginary parts.
Simplified versions of the
scheme have been introduced in \citere{imflscheme}.
\end{enumerate}

Tables~\ref{tab:cc10a} and \ref{tab:cc10b} contain some results%
\footnote{The input data for the two tables are
different, because the complex-mass scheme \cite{de98b} requires a minimal 
set of input parameters with relation \refeq{eq:csw},
while $\sw$ was deduced from the Fermi
constant in the evaluation of the fermion-loop scheme in \citere{flscheme}.}
on the cross section for the ``CC10 process'' 
$\Pep\Pem\to\Pu\bar\Pd\mu^-\bar\nu_\mu$ for the different finite-width
schemes. 
\begin{table}
\centerline{
\begin{tabular}[t]{|c|c|c|c|c|}
\hline
$\sqrt{s}/\GeV$ & 200 & 500 & 1000 & 2000 \\
\hline\hline
Fixed width   & 712.8(2) & 237.3(1) & 60.34(4) & 13.97(1)
\\ \hline
Running width & 712.7(2) & 238.7(1) & 65.74(4) & 34.40(2)
\\ \hline
Complex mass  & 712.4(2) & 237.1(1) & 60.31(4) & 13.97(1)
\\ \hline
\end{tabular}
}
\caption{CC10 cross section in fb for various finite-width schemes 
(based on the results of \protect\citere{de98b}).}
\label{tab:cc10a}
\centerline{
\begin{tabular}[t]{|c|c|c|c|c|}
\hline
$\sqrt{s}/\GeV$ & 200 & 500 & 1000 & 2000 \\
\hline\hline
Fixed width    & 673.08(4) & 224.05(3) & 56.90(1) & 13.19(1)
\\ \hline
Running width  & 672.96(3) & 225.45(3) & 62.17(1) & 33.06(1)
\\ \hline
Full fermion loops & 683.7(1)  & 227.9(2)  & 58.0(1)  & 13.57(4)
\\ \hline
Imag.\ fermion loops & 673.1(1)  & 224.5(7)  & 56.8(1)  & 13.18(4)
\\ \hline
\end{tabular}
}
\caption{CC10 cross section in fb for various finite-width schemes 
(taken from \protect\citere{flscheme}).
The two versions for the fermion-loop scheme (numbers generated by 
{\sl WTO} \protect\cite{WTO}) differ by the real parts of the fermion loops,
which are included in the upper row but not in the lower.}
\label{tab:cc10b}
\end{table}
The energies are typical of LEP2 and future linear colliders,
and the phase space is restricted by the ``canonical LEP2 cuts''
of \citere{lep2repWevgen}. 
In the high-energy limit, SU(2) gauge invariance implies delicate
cancellations between different diagrams. The complex-mass and the
fermion-loop schemes fully respect these cancellations, since the
underlying Ward identities hold, resulting in reliable predictions 
for high energies. Both the fixed and running width break SU(2)
invariance so that the gauge cancellations are disturbed. While the
gauge-invariance-breaking terms grow like $s/\MW^2$ for the running
width scheme, leading to totally wrong results already in the TeV range,
those terms are suppressed for the fixed-width scheme (see above), 
still yielding reasonable results for high energies.

\subsection{Looking beyond lowest order}

Among the methods to introduce finite gauge-boson widths in
tree-level amplitudes, the field-theoretically most convincing one is
the fermion-loop scheme, since the widths are consistently generated by
the inclusion of the relevant higher-order effects. It is therefore
natural to worry about a generalization of this scheme when considering
higher-order corrections to amplitudes. There are basically two sources
of limitation for the fermion-loop scheme. 

Firstly, the fermion-loop scheme is not applicable in the presence of 
resonant particles that do not exclusively decay into fermions. For such
particles, parts of the decay width are contained in bosonic
corrections. The Dyson summation of fermionic {\it and}
bosonic ${\cal O}(\alpha)$ corrections leads to inconsistencies in the
usual field-theoretical approach, i.e.\ Ward identities
are broken in general. This is due to the fact that the
bosonic ${\cal O}(\alpha)$ contributions to vertex functions do not obey 
the ``naive Ward identities''. The problem is circumvented by employing
the background-field formalism \cite{de95}, in which these naive
identities are valid. This implies \cite{de96} that a consistent Dyson
summation of fermionic and bosonic corrections to any order in $\alpha$
does not disturb Ward identities.
Therefore, the background-field approach provides a natural generalization
of the fermion-loop scheme.
We recall that any resummation formalism goes beyond a strict
order-by-order calculation and necessarily involves ambiguities
in relative order $\alpha^n$ if not all $n$-loop diagrams are included. 
This kind of scheme dependence, which in particular concerns gauge 
dependences, is only resolved by successively calculating the missing
orders.

Secondly, the consistent resummation of all 
${\cal O}(\alpha)$ loop corrections does not automatically lead to 
${\cal O}(\alpha)$ precision in the predictions if resonances are involved.
The imaginary parts of one-loop self-energies generate only
tree-level decay widths so that directly on resonance one order in
$\alpha$ is lost. To obtain also full ${\cal O}(\alpha)$
precision in these cases, the imaginary parts of the two-loop
self-energies are required. However, how and whether this two-loop
contribution can be included in a practical way without violating 
Ward identities is still an open problem. 
Taking the imaginary parts of all two-loop contributions solves the 
problem in principle, at least for the background-field approach, but 
this is certainly impractical.

\section{Electroweak radiative corrections}

\subsection{Relevance of electroweak corrections}

Present-day Monte Carlo generators
for off-shell W-pair production (see e.g.\ \citere{lep2repWevgen})
typically include only universal electroweak ${\cal O}(\alpha)$ corrections%
\footnote{The QCD corrections for hadronic final states are discussed 
in \citere{qcdcorr}.},
such as the running of the electromagnetic coupling, $\alpha(q^2)$, 
leading corrections entering via the $\rho$-parameter,
the Coulomb singularity \cite{coul}, which is important near threshold,
and mass-singular logarithms 
$\alpha\ln(\Me^2/Q^2)$ from initial-state radiation. 
In leading order, the scale $Q^2$ is not determined and has to be set to
a typical scale for the process; for the following we take $Q^2=s$.

Since the full ${\cal O}(\alpha)$ correction is not known for off-shell
W~pairs, the size of the neglected ${\cal O}(\alpha)$ contributions is
estimated by inspecting on-shell W-pair production, for which the exact 
${\cal O}(\alpha)$ correction and the leading contributions were
given in \citeres{vrceeww} and \cite{bo92}, respectively.
Table \ref{tab:nlRCs} shows the difference between an ``improved
Born approximation'' $\de_{\mathrm{IBA}}$, which is based on the 
above-mentioned universal corrections, and the corresponding full
${\cal O}(\alpha)$ correction $\de$ to the Born cross-section integrated over
the W-production angle $\theta$ for some centre-of-mass energies $\sqrt{s}$.
\begin{table}
\centerline{
\begin{tabular}{|c||c||c|c|c|c|c|c|}
\hline
$\theta$ range & $\sqrt{s}/\GeV$ &
161 & 175 & 200 & 500 & 1000 & 2000 \\
\hline\hline
$0^\circ$$<$$\theta$$<$$180^\circ$ & 
$(\de_{\mathrm{IBA}}-\de)/\%$ 
& 1.5 & 1.3 & 1.5 & 3.7 & 6.0 & 9.3 \\
\cline{1-1} \cline{3-8}
$10^\circ$$<$$\theta$$<$$170^\circ$ &&
1.5 & 1.3 & 1.5 & 4.7 & 11 & 22 \\
\hline
\end{tabular}
}
\caption{Size of ``non-leading'' corrections to on-shell W-pair
production ($\de_{\mathrm{IBA}}$ and $\de$ include only soft-photon
emission).}
\label{tab:nlRCs}
\end{table}
More details and results can be found in \citeres{lep2repWcs,crad96}.
The quantity $\de_{\mathrm{IBA}}-\de$ corresponds to the neglected
non-leading corrections and amounts to $\sim 1$--2\% for LEP2 energies,
but to $\sim 10$--20\% in the TeV range. Thus, in view of the needed
$0.5$\% level of accuracy for LEP2 and all the more for energies of
future linear colliders, the inclusion of non-leading corrections is
indispensable. 
The large contributions in $\de_{\mathrm{IBA}}-\de$ at high energies are
due to terms such as $\alpha\ln^2(s/\MW^2)$, which arise from vertex 
and box corrections and can be read off from the high-energy 
expansion \cite{be93} of the virtual and soft-photonic ${\cal O}(\alpha)$ 
corrections.

A first step of including electroweak corrections in an event 
generator was made by Jadach et al.\ \cite{ja97}, who took into account the
full virtual \cite{vrceeww} and real-photonic \cite{brrceeww}
${\cal O}(\alpha)$ corrections to the on-shell W-pair production process,
but neglected the corrections associated with the decay of the W~bosons.
In order to gain an accuracy of 0.5--1\% for realistic
observables, it is, however, necessary to develop a more complete
strategy for the full process $\Pep\Pem\to\PW\PW\to 4f$. 

\subsection{Double-pole approximation}
\label{se:DPA}

Fortunately, the full off-shell calculation for the process 
$\Pep\Pem\to\PW\PW\to 4f$ in ${\cal O}(\alpha)$ is not needed for
most applications. Sufficiently above the W-pair threshold 
a good approximation should be obtained by taking into account
only the doubly-resonant part of the amplitude \refeq{eq:Mstruc}, 
leading to an error of the order of $\alpha\Ga_\PW/(\pi\MW)\lsim 0.1\%$. 
In such a ``pole scheme'' calculation \cite{ae94,st91}, also called
double-pole approximation (DPA), the numerator 
$R_{+-}(k_+^2,k_-^2)$ has to be replaced by the gauge-independent residue 
$R_{+-}(\MW^2,\MW^2)$. 

Doubly-resonant corrections to 
$\Pep\Pem\to\PW\PW\to 4f$ can be classified into two types
\cite{lep2repWcs,be94,ae94}:
factorizable and non-factorizable corrections.
The former are those that correspond either to W-pair production 
or to W~decay. They are represented by the schematic diagram of 
\reffi{fig:fRCsdiag}, in which the shaded blobs contain all one-loop
corrections to the production and decay processes, and the open blobs
include the corrections to the W~propagators. 
\begin{figure}
\centerline{
\begin{picture}(155,85)(0,0)
\SetScale{.8}
\ArrowLine(30,50)( 5, 95)
\ArrowLine( 5, 5)(30, 50)
\Photon(30,50)(150,80){2}{11}
\Photon(30,50)(150,20){2}{11}
\ArrowLine(150,80)(190, 95)
\ArrowLine(190,65)(150,80)
\ArrowLine(190, 5)(150,20)
\ArrowLine(150,20)(190,35)
\GCirc(30,50){10}{.5}
\GCirc(90,65){10}{1}
\GCirc(90,35){10}{1}
\GCirc(150,80){10}{.5}
\GCirc(150,20){10}{.5}
\DashLine( 70,0)( 70,100){2}
\DashLine(110,0)(110,100){2}
\put(40,21){W}
\put(40,53){W}
\put(95, 8){W}
\put(95,67){W}
\put(-12, 5){$\Pem$}
\put(-12,70){$\Pep$}
\put(160, 1){$\bar f_4$}
\put(160,24){$f_3$}
\put(160,50){$\bar f_2$}
\put(160,75){$f_1$}
\put(-25,-10){\footnotesize On-shell production}
\put(100,-10){\footnotesize On-shell decays}
\SetScale{1}
\end{picture}
}
\caption{Diagrammatic structure of factorizable corrections to
$\Pep\Pem\to\PW\PW\to 4f$.}
\label{fig:fRCsdiag}
\end{figure}
The remaining corrections
are called non-factorizable, since they do not contain the product of
two independent Breit--Wigner-type resonances for the W~bosons, i.e.\
the production and decay subprocesses are not independent in this case. 
Non-factorizable corrections include all diagrams involving particle
exchange between these subprocesses. Simple power-counting arguments
reveal that such diagrams only lead to doubly-resonant contributions if
the exchanged particle is a photon with energy
$E_\gamma\lsim\Gamma_\PW$; all other non-factorizable diagrams 
are negligible in DPA. Two relevant diagrams are shown in
\reffi{fig:nfRCsdiags}, where the full blobs represent tree-level
subgraphs.
\begin{figure}
\centerline{
\begin{picture}(110,75)(0,8)
\SetScale{0.8}
\ArrowLine(30,50)( 5, 95)
\ArrowLine( 5, 5)(30, 50)
\Photon(30,50)(90,80){2}{6}
\Photon(30,50)(90,20){2}{6}
\GCirc(30,50){10}{0}
\Vertex(90,80){1.2}
\Vertex(90,20){1.2}
\ArrowLine(90,80)(120, 95)
\ArrowLine(120,65)(105,72.5)
\ArrowLine(105,72.5)(90,80)
\Vertex(105,72.5){1.2}
\ArrowLine(120, 5)( 90,20)
\ArrowLine( 90,20)(105,27.5)
\ArrowLine(105,27.5)(120,35)
\Vertex(105,27.5){1.2}
\Photon(105,27.5)(105,72.5){2}{4.5}
\put(89,40){$\gamma$}
\put(42,60){$W$}
\put(42,15){$W$}
\SetScale{1}
\end{picture}
\begin{picture}(190,75)(0,8)
\SetScale{.8}
\ArrowLine(30,50)( 5, 95)
\ArrowLine( 5, 5)(30, 50)
\Photon(30,50)(90,80){2}{6}
\Photon(30,50)(90,20){2}{6}
\GCirc(30,50){10}{0}
\Vertex(90,80){1.2}
\Vertex(90,20){1.2}
\ArrowLine(90,80)(120, 95)
\ArrowLine(120,65)(105,72.5)
\ArrowLine(105,72.5)(90,80)
\ArrowLine(120, 5)( 90,20)
\ArrowLine( 90,20)(120,35)
\Vertex(105,72.5){1.2}
\PhotonArc(120,65)(15,150,270){2}{3}
\put(42,60){W}
\put(42,15){W}
\put(75,40){$\gamma$}
\DashLine(120,0)(120,100){6}
\PhotonArc(120,35)(15,-30,90){2}{3}
\Vertex(135,27.5){1.2}
\ArrowLine(150,80)(120,95)
\ArrowLine(120,65)(150,80)
\ArrowLine(120, 5)(150,20)
\ArrowLine(150,20)(135,27.5)
\ArrowLine(135,27.5)(120,35)
\Vertex(150,80){1.2}
\Vertex(150,20){1.2}
\Photon(210,50)(150,80){2}{6}
\Photon(210,50)(150,20){2}{6}
\ArrowLine(210,50)(235,95)
\ArrowLine(235, 5)(210,50)
\GCirc(210,50){10}{0}
\put(140,60){W}
\put(140,15){W}
\SetScale{1}
\end{picture}
}
\caption{Examples of virtual and real non-factorizable 
corrections to $\Pep\Pem\to\PW\PW\to 4f$.}
\label{fig:nfRCsdiags}
\vspace*{1.2em}
\centerline{
\begin{picture}(100,70)(0,7)
\SetScale{.8}
\ArrowLine(30,50)( 5, 95)
\ArrowLine( 5, 5)(30, 50)
\Photon(30,50)(90,80){2}{6}
\Photon(30,50)(90,20){2}{6}
\Photon(70,30)(70,70){2}{4}
\Vertex(70,30){1.2}
\Vertex(70,70){1.2}
\GCirc(30,50){10}{0}
\Vertex(90,80){1.2}
\Vertex(90,20){1.2}
\ArrowLine(90,80)(120, 95)
\ArrowLine(120,65)(90,80)
\ArrowLine(120, 5)( 90,20)
\ArrowLine( 90,20)(120,35)
\put(63,40){$\gamma$}
\put(37,57){W}
\put(37,17){W}
\SetScale{1}
\end{picture} }
\caption{Feynman graph contributing to both factorizable and
non-factorizable corrections.}
\label{fig:nfRCsdiag2}
\end{figure}
We note that diagrams involving photon exchange between the W~bosons
(see \reffi{fig:nfRCsdiag2}) contribute both to factorizable and
non-factorizable corrections; otherwise the splitting into those parts
is not gauge-invariant.
The non-factorizable corrections to $\Pep\Pem\to\PW\PW\to 4f$ will be
discussed in \refse{se:nfRCs} in more detail.

The factorizable corrections consist of contributions from virtual 
corrections and real-photon bremsstrahlung. The known results on
the virtual corrections to the pair production \cite{vrceeww}
and the decay \cite{rcwdecay} of on-shell W~bosons can be used as building
blocks for the DPA. Because of the complex structure of the virtual 
corrections to the production process, simple approximations are desirable.
In the LEP2 energy range, approximations \cite{bo92,sc97} that are
mainly based on universal corrections describe the total cross section
within $\sim 0.5\%$ and the angular distribution within 1--2\% relative to the 
lowest order. For energies above $500\GeV$, a consistent high-energy 
expansion \cite{be93,sc98} reaches per-cent accuracy whenever the cross 
section is sizeable. However, a satisfying, simple approximation for all 
relevant energies is not available.

The formulation of a consistent DPA for the bremsstrahlung
corrections is non-trivial. The main complication originates from
the emission of photons from the resonant W~bosons. A radiating
W~boson involves two propagators whose momenta differ by the
momentum of the emitted photon. If the photon momentum is large 
($E_\gamma\gg\Gamma_\PW$), the resonances of these two propagators are
well separated in phase space, and their contributions can be associated
with photon radiation from exactly one of the production or decay
subprocesses. For soft photons ($E_\gamma\ll\Gamma_\PW$) a similar
splitting is possible. However, for $E_\gamma\sim\Gamma_\PW$ the two
resonance factors for the radiating W~boson overlap so that a simple
decomposition into contributions associated with the subprocesses is not
obvious.

A full application of the DPA to $\Pep\Pem\to\PW\PW\to 4f$ has not yet 
been presented in the literature, but first preliminary results have
been shown by Berends \cite{fb98} at this conference.

\subsection{Photon radiation and W line shape}
\label{se:Wlineshape}

A thorough description of real-photon emission is of particular
importance for the realistic prediction of the W~line shape, which
is the basic observable for the reconstruction of the W-boson mass from
the W-decay products. This fact can be easily understood by comparing the
impact of photon radiation on the line shape of the W~boson with the one
of the Z~boson, observed in $\Pep\Pem\to\PZ\to f\bar f$ at LEP1 and the
SLC (see \reffi{fig:lineshape}). 
\begin{figure}
\hspace*{2em}
\begin{picture}(130,90)(0,10)
\ArrowLine(30,50)( 5, 80)
\ArrowLine( 5,20)(30, 50)
\Photon(30,50)(90,50){2}{6}
\Photon(30,50)(38,80){2}{4}
\Photon(90,50)(98,80){2}{4}
\ArrowLine( 90,50)(115,80)
\ArrowLine(115,20)( 90,50)
\GCirc(30,50){8}{0}
\GCirc(90,50){8}{0}
\put(55,35){Z}
\put(28,82){{$\gamma$}}
\put(88,82){{$\gamma$}}
\put(-12,15){$\Pem$}
\put(-12,75){$\Pep$}
\put(120,15){$\bar f$}
\put(120,75){$f$}
\end{picture}
\hspace*{3em}
\begin{picture}(130,90)(0,10)
\ArrowLine(30,50)( 5, 80)
\ArrowLine( 5,20)(30, 50)
\Photon(30,50)(90,70){-2}{6}
\Photon(30,50)(90,30){2}{6}
\Photon(30,50)(38,80){2}{4}
\Photon(90,70)(105,95){2}{4}
\Photon(90,30)(105, 5){2}{4}
\Photon(60,60)( 75,85){2}{4}
\Photon(60,40)( 75,15){2}{4}
\ArrowLine( 90,70)(120,85)
\ArrowLine(120,55)( 90,70)
\ArrowLine(120,15)( 90,30)
\ArrowLine( 90,30)(120,45)
\Vertex(60,60){1.2}
\Vertex(60,40){1.2}
\GCirc(30,50){8}{0}
\GCirc(90,70){8}{0}
\GCirc(90,30){8}{0}
\put(43,26){W}
\put(43,63){W}
\put(28,82){{$\gamma$}}
\put(65,90){{$\gamma$}}
\put(90,94){{$\gamma$}}
\put(60,15){{$\gamma$}}
\put(90, 5){{$\gamma$}}
\put(-12,15){$\Pem$}
\put(-12,75){$\Pep$}
\put(125,11){$\bar f_4$}
\put(125,39){$f_3$}
\put(125,55){$\bar f_2$}
\put(125,85){$f_1$}
\end{picture}
\caption{Illustration of photon radiation in $\Pep\Pem\to\PZ\to f\bar f$
and $\Pep\Pem\to\PW\PW\to 4f$.}
\label{fig:lineshape}
\end{figure}

The Z~line shape is defined as a function of $s$, which is fully 
determined by the initial state, by the cross section $\sigma(s)$.
Photon radiation from the initial state effectively reduces the value of 
$s$ that is ``seen'' by the Z boson so that $\sigma(s)$ also receives
resonant contributions for $s>\MZ^2$, induced by this {\it radiative
return} to the Z~resonance and known as {\it radiative tail}. Final-state 
radiation is not enhanced by
such kinematical effects, thus yielding moderate corrections.

The W~line shape is reconstructed from the kinematical variables in the
final state. More precisely, it is defined by the distributions
$\rd\sigma/\rd M_\pm^2$, where $M_\pm^2$ are the reconstructed invariant
masses of the $\PW^\pm$~bosons. We now consider the fermion pair 
$f_1(k_1) \bar f_2(k_2)$ produced by a nearly resonant W~boson with momentum
$k_+$, i.e.\ $k_+^2\sim\MW^2$.
In this case, photon radiation from the final state decreases the
invariant mass of this fermion pair, i.e.\ 
$(k_1+k_2)^2<k_+^2=(k_1+k_2+k_\gamma)^2$,
while initial-state radiation leads to 
$(k_1+k_2)^2=k_+^2<(k_1+k_2+k_\gamma)^2$.
Thus, a consistent identification of $M_+^2=(k_1+k_2)^2$ also leads to a
radiative tail, but now induced by final-state radiation and for 
$M_+^2<\MW^2$. However, such an identification is experimentally not
possible for almost all cases%
\footnote{Semi-leptonic final states with a muon may be an exception,
where $(k_\mu+k_{\nu_\mu})^2$ could be determined from all detected
final-state particles other than the muon.},
since nearly collinear and soft photons in the final state cannot be fully
separated from the outgoing fermions (except for muons). A realistic
definition of $M_\pm^2$ necessarily depends on the details of the
experimental treatment of photons in the final state, underlining the
importance of a careful investigation of the W~line shape in the
presence of photon radiation.

The line-shape distortion by photon emission was discussed in \citere{be98} 
also quantitatively. Based on an exact treatment of the Z~boson line shape in
$\nu_\mu\bar\nu_\mu\to\PZ\PZ\to\Pem\Pep\nu_\tau\bar\nu_\tau$, an
estimate for the process $\Pep\Pem\to\PW\PW\to 4f$ could be obtained by
an analysis in leading-logarithmic accuracy. Assuming the idealized
definition $M_+^2=(k_1+k_2)^2$, the authors of \citere{be98} find shifts
in the peak position of $\rd\sigma/\rd M_+^2$ of several times $-10\MeV$
and peak reduction factors in the range 0.95--0.75. The maximal effect
was obtained for $\PWp\to\Pep\nu_\Pe$, leading to 
$\Delta M_+ \sim -45\MeV$. Note, however, that these large corrections
are formally due to mass-singular logarithms such as
$\alpha\ln(\Me/\MW)$, which are effectively replaced by logarithms of a
minimum opening angle for collinear photon emission in more realistic
definitions of $M_+^2$. Therefore, it is expected that the corrections
are somewhat weakened in realistic situations.

\subsection{The process $\Pep\Pem\to 4f+\gamma$}

The process $\Pep\Pem\to 4f+\gamma$ does not only yield important 
corrections to $\Pep\Pem\to 4f$, as explained above, it is also
interesting in its own right, since it involves both triple and quartic 
gauge-boson couplings. 
Most of the existing work on hard-photon radiation in W-pair production
is based on the approximation of stable W~bosons (see \citere{brrceeww}
and references in \citeres{lep2repWcs,de98b}). 
A first step of including the off-shellness of W~bosons in
$\Pep\Pem\to\PW\PW\to 4f+\gamma$ was done in \citere{ae91}, where only
photon emission from the signal diagrams of W-pair production was taken
into account. 
However, it is desirable to have a full lowest-order calculation
for $\Pep\Pem\to 4f+\gamma$ for two reasons. As described above, the
definition of the DPA for $\Pep\Pem\to\PW\PW\to 4f+\gamma$ is
non-trivial so that possible versions of DPAs should be carefully
compared to the full result. Secondly, one expects a similar impact of
off-shell effects, for instance induced by background diagrams, on
$\Pep\Pem\to 4f+\gamma$ as in the case without photon, where such
effects can reach a significant fraction of the full cross section. 
Therefore, they should be included at least in predictions for 
detectable photons, which define their own class of processes.
The first full lowest-order calculation of $\Pep\Pem\to 4f+\gamma$ was
performed in \citere{fu94} for a definite final state, and a fully 
numerical treatment of all possible final states was presented in
\citere{ca97}. In the following we focus on the recent calculation of
\citere{de98b}, where an event generator for all final states 
$4f+\gamma$ is described. 

\begin{figure}
\centerline{
{ \setlength{\unitlength}{1cm}
\begin{picture}(11,5.0)
\put(2.0,-0.2){\footnotesize $E_\gamma/\GeV$}
\put(7.8,-0.2){\footnotesize $E_\gamma/\GeV$}
\put( 0.6,1.0){\footnotesize $\sqrt{s}=190\GeV$}
\put( 6.1,1.0){\footnotesize $\sqrt{s}=500\GeV$}
\put(1.2,0.8){\includegraphics{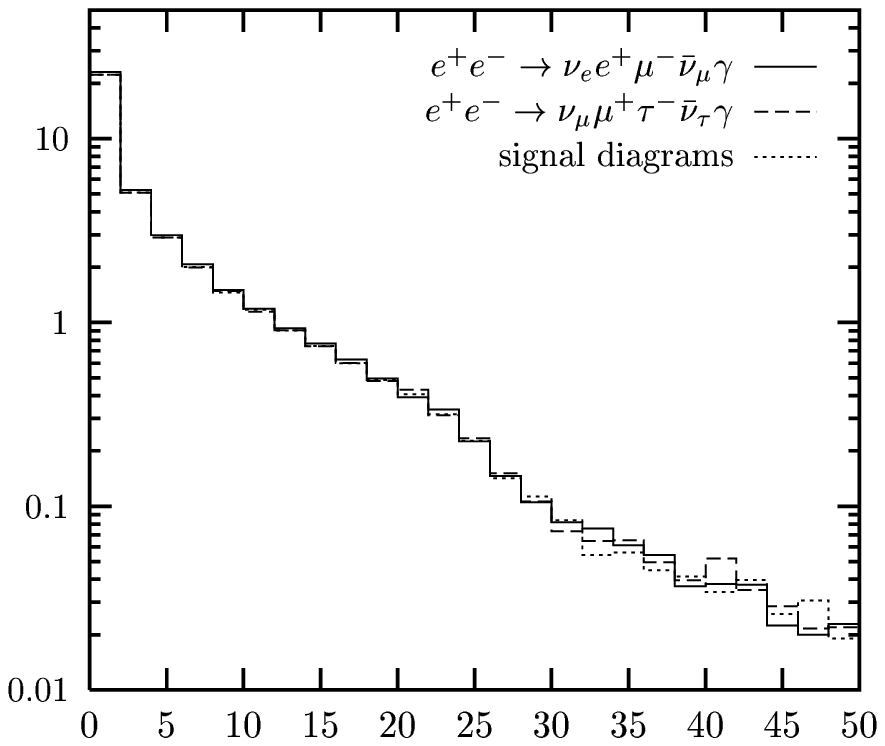}}
\put(6.7,0.8){\includegraphics{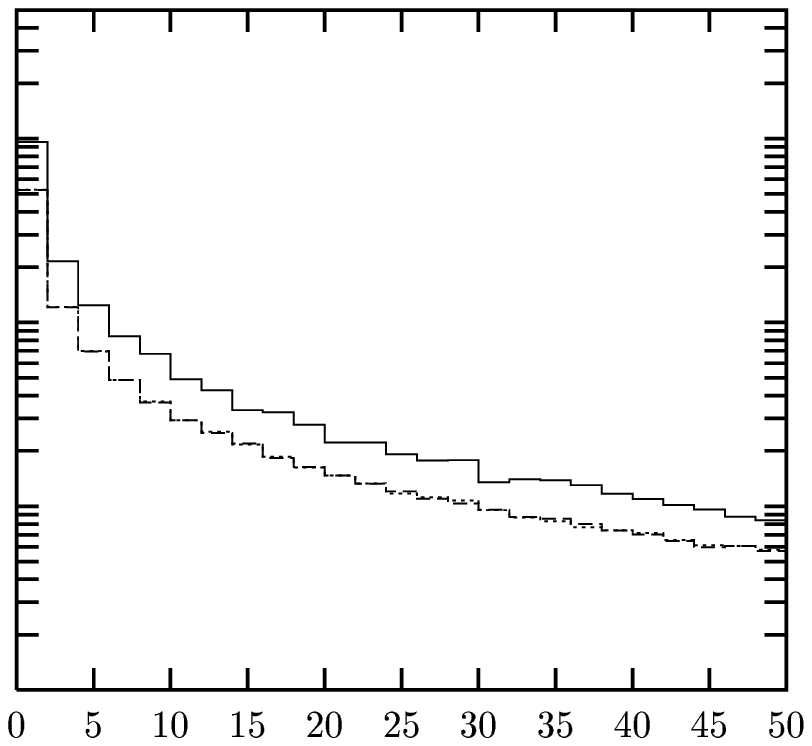}}
\end{picture} }
}
\caption{The photon-energy spectrum 
$(\rd\sigma/\rd E_\gamma) / (\mathrm{fb/GeV})$
of $\Pep\Pem\to 4f+\gamma$ for two
different leptonic final states and for the approximation of including
only photon emission from W-pair signal diagrams 
(based on the results of \protect\citere{de98b}).}
\label{fig:ee4fa}
\end{figure}

In the event generator of \citere{de98b} the matrix elements are
constructed from simple generic functions, similar to the approach
pursued in {\sl Excalibur} \cite{Be94} for $4f$ final states. Moreover,
different schemes for treating gauge-boson widths are implemented, 
namely the ones for constant and running widths, as well as the
complex-mass scheme (see \refse{se:widths}).
Table~\ref{tab:ee4fa} contains some sample results on the total cross
section (applying the canonical cuts of \citere{lep2repWevgen}) for a 
final state of four leptons and a photon, evaluated for LEP2 and 
linear-collider energies with different finite-width treatments.
\begin{table}
\centerline{
\begin{tabular}{|c|c|c|c|c|}
\hline
$\sqrt{s}/\GeV$ & 200 & 500 & 1000 & 2000 \\
\hline\hline
Fixed width   & 82.3(3) & 26.3(1) & 7.29(2) & 2.17(1) \\
\hline
Running width & 82.5(3) & 26.8(1) & 8.29(2) & 6.29(2) \\
\hline
Complex mass  & 82.0(3) & 26.4(1) & 7.26(2) & 2.17(1) \\
\hline
\end{tabular}
}
\caption{Cross section for
$\Pep\Pem\to\nu_\mu \mu^+\tau^-\bar\nu_\tau + \gamma$
in fb for various finite-width schemes 
(based on the results of \protect\citere{de98b}).}
\label{tab:ee4fa}
\end{table}
Similar to the case without photon emission, the SU(2)-breaking effects
induced by a running width render the predictions totally wrong in the
TeV range. For a constant width such effects are suppressed, as can be
seen from a comparison with the results of the complex-mass scheme,
which exactly preserves gauge invariance.
The influence of background diagrams on predictions for 
$\Pep\Pem\to\PW\PW\to 4f+\gamma$ is illustrated in \reffi{fig:ee4fa},
where the photon-energy distribution is shown for the final states
$\nu_\Pe \Pep\mu^-\bar\nu_\mu\gamma$ and
$\nu_\mu \mu^+\tau^-\bar\nu_\tau\gamma$. The prediction that is based
on photon emission from signal diagrams only is also included for
comparison.
While the impact of background diagrams is small at the LEP2 energy of
$200\GeV$, there is a large background contribution already at $500\GeV$
for final states with $\Pepm$. The main effect is due to
forward-scattered $\Pepm$, which is familiar
from the results on $\Pep\Pem\to 4f$. More numerical results on
$\Pep\Pem\to 4f+\gamma$ can be found in \citere{de98b}.
\looseness -1

The results on $\Pep\Pem\to 4f+\gamma$ with detectable photons can be used 
as building block for four-fermion production with and without photon
emission, $\Pep\Pem\to\PW\PW\to 4f(+\gamma)$.
To this end, contributions of soft and collinear photon emission as well 
as virtual corrections have to be added.
If a DPA is applied, care has to be taken to avoid mismatch 
between IR and mass singularities in virtual and real corrections. 
The handling of such contributions strongly depends on the details of the
adopted approximations, i.e.\ on how the off-shellness of the W~bosons is
introduced, and whether background diagrams are included.
Moreover, one has to make sure to avoid double-counting of real
corrections when adding the non-factorizable corrections, which are
described next.
\looseness -1

\subsection{Non-factorizable corrections}
\label{se:nfRCs}

As already explained in \refse{se:DPA}, non-factorizable corrections 
account for the exchange of photons with $E_\gamma\lsim\Gamma_\PW$
between the W-pair production and W~decay subprocesses
(see \reffis{fig:nfRCsdiags},\ref{fig:nfRCsdiag2}).
Already before their explicit calculation, it was shown
\cite{fa94} that such corrections 
vanish if the invariant masses of both W~bosons are integrated over. 
Thus, they do not influence pure angular distributions,
which are of particular importance for the analysis of gauge-boson
couplings. For exclusive quantities the non-factorizable corrections are 
non-vanishing. A first hint on their actual size was obtained by 
investigating the non-factorizable correction that is contained in the 
Coulomb singularity \cite{kh95}.

The explicit analytical calculation of the non-factorizable corrections was 
performed by different groups \cite{me96,be97,de98a}%
\footnote{The original result of the older calculation \cite{me96} does
not agree with the two more recent results \cite{be97,de98a}, which are
in mutual agreement. As known 
from the authors of \citere{me96}, 
their corrected results also agree with the ones of \citeres{be97,de98a}.}.
In these studies, the photon momentum was integrated over, resulting in
a correction factor to the differential Born cross section for the process
without photon emission. This correction factor is non-universal \cite{de98a} 
in the sense that it depends on the
parametrization of phase space. This is due to the problem of defining
the invariant masses of the W~bosons in the presence of real photon
emission, which has already been described in \refse{se:Wlineshape}.
In \citeres{me96,be97,de98a} the invariant masses $M_\pm$ of the \PWpm~bosons
are identified with the invariant masses of the corresponding final-state
fermion pairs. The analytical results show that
all effects from the initial $\Pep\Pem$ state
cancel so that the correction factor does not depend on the
W-production angle and is also applicable to processes such as
$\gamma\gamma\to\PW\PW\to 4f$. Fermion-mass singularities appear in
individual contributions, but cancel in the sum of virtual and real
corrections. Moreover, the correction factor vanishes
like $(M_\pm^2-\MW^2)/(\Gamma_\PW\MW)$ on resonance and tends to zero in
the high-energy limit, both leading to a suppression of the non-factorizable
corrections with respect to the factorizable ones.

\begin{figure}
\centerline{
\setlength{\unitlength}{1cm}
\begin{picture}(11,5)
\put(0,0){\includegraphics{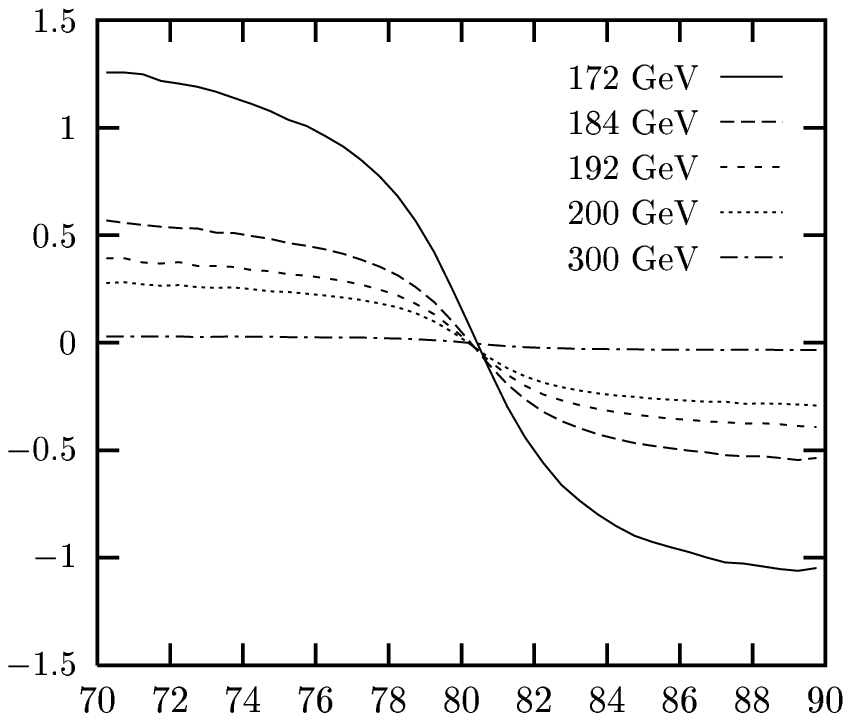}}
\put(0,0){\includegraphics{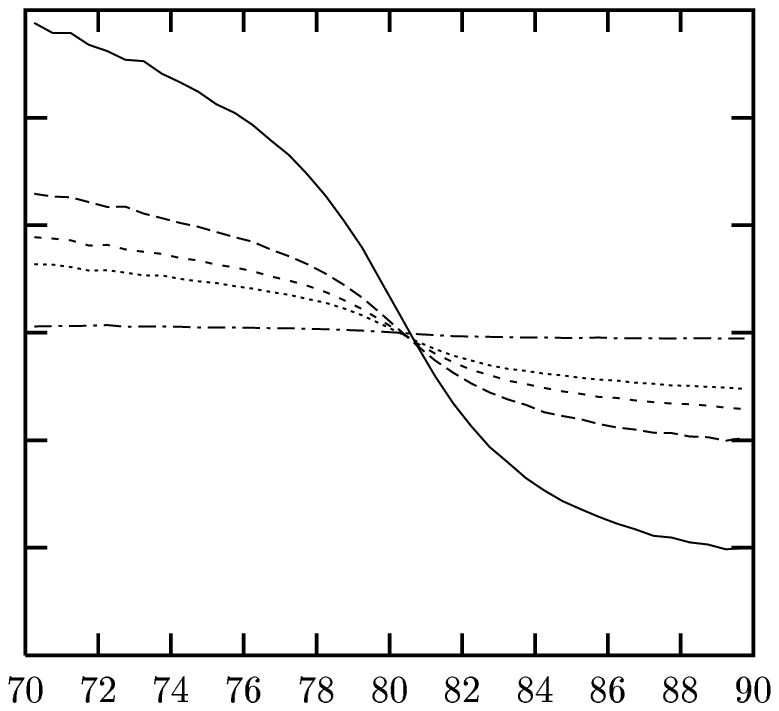}}
\put(-0.4,3.9){\makebox(1,1)[c]{\small$\de_{\mathrm{nf}}/\%$}}
\put(2.5,-0.6){\makebox(1,1)[cc]{{\small$M_\pm/{\GeV}$}}}
\put(7.8,-0.6){\makebox(1,1)[cc]{{\small$M_\pm/{\GeV}$}}}
\put(1.4,1.1){\footnotesize leptons}
\put(6.7,1.1){\footnotesize hadrons}
\end{picture}
}
\caption{Relative non-factorizable corrections to single-invariant-mass
distributions for $\Pep\Pem\to\PW\PW\to 4f$ with purely leptonic 
and hadronic final states (taken from \protect\citere{de98a}).}
\label{fig:nonfactnum}
\end{figure}

The non-factorizable corrections to $\Pep\Pem\to\PW\PW\to 4\,$leptons were 
numerically evaluated in \citere{be97} and for all final states in 
\citere{de98a}.
The corrections to single-invariant-mass distributions 
(see \reffi{fig:nonfactnum}) turn out to be
qualitatively similar for all final states and are of the order
of $\sim 1\%$ for LEP2 energies, shifting the maximum of the distributions
by 1--2$\MeV$, which is small with respect to 
LEP2 accuracy \cite{lep2repWmass}.
Multiple distributions in angular or energy variables and in at least
one of the invariant masses of the W~bosons receive larger corrections,
of a few per cent.

The non-factorizable corrections to $\Pep\Pem\to\PZ\PZ\to 4f$ are
discussed in \citere{de98a}, too. In this case, the corrections to the
invariant-mass distributions are further suppressed and of the order of
$\sim 0.1\%$, which is phenomenologically negligible. The suppression is
due to an antisymmetry of the dominant parts of the differential cross
sections in the angular integrations. Combined angular and
invariant-mass distributions that break this antisymmetry in the 
integration get large corrections of several per cent. However, since the 
cross section for Z-pair production is, by an order of magnitude, smaller
than the one for W~pairs, these corrections are still not significant at
LEP2. 

Although non-factorizable corrections to four-fermion production turn out 
to be small with respect to LEP2 accuracy, they
can be of relevance at future $\Pep\Pem$ colliders with higher
luminosity.

\section*{Acknowledgements}
I would like to thank the organizers for their kind invitation and for 
providing a very pleasant atmosphere during the conference.
F.A.~Berends, A.~Denner, M.~Gr\"unewald, W.~Hollik, T.~Riemann, M.~Roth, 
D.~Schildknecht, and D.~Wackeroth are
gratefully acknowledged for helpful discussions.

\end{document}